\documentclass[conference]{IEEEtran}
\IEEEoverridecommandlockouts
\usepackage{cite}
\usepackage{amsmath,amssymb,amsfonts}
\usepackage{algorithmic}
\usepackage{graphicx}
\usepackage{textcomp}
\usepackage{xcolor}
\usepackage[nolist]{acronym}
\usepackage{paralist}
\usepackage{booktabs}
\def\BibTeX{{\rm B\kern-.05em{\sc i\kern-.025em b}\kern-.08em
    T\kern-.1667em\lower.7ex\hbox{E}\kern-.125emX}}

\begin{document}


\begin{acronym}
    \acro{BMWK}{German Federal Ministry for Economic Affairs and Climate Action}
    \acro{DSO}{Distribution System Operator}
    \acro{ICT}{Information and Communications Technology}
    \acro{IDS}{Intrusion Detection System}
    \acro{IT}{Information Technology}
    \acro{OT}{Operational Technology}
    \acro{ET}{Energy Technology}
    \acro{EMS}{Energy Management System}
    \acro{RTU}{Remote Terminal Unit}
    \acro{SCADA}{Supervisory Control and Data Acquisition}
    \acro{SMGW}{Smart Meter Gateway}
    \acro{TSO}{Transmission System Operator}
    \acro{SGAM}{Smart Grid Architecture Model}
    \acro{PCAP}{Packet Capture}
    \acro{vRTU}{virtual Remote Terminal Unit}
    \acro{MTU}{Master Terminal Unit}
    \acro{DER}{Decentralized Energy Resource}
    \acro{VED}{Virtual Edge Devices}
    \acro{DMZ}{Demilitarized Zone}
    \acro{IED}{Intelligent Electronic Device}
    \acro{VPN}{Virtual Private Network}
    \acro{VPP}{Virtual Power Plant}
    \acro{VED}{Virtual Edge Device}
    \acro{SG}{Smart Grid}
    \acro{CVE}{Common Vulnerabilities and Exposures}
    \acro{RCE}{Remote Code Execution}
    \acro{PE}{Privilege Escalation}
    \acro{SAM}{Simulated Attacker Model}
    \acro{SUID}{set-user-ID}
    \acro{FDI}{False Data Injection}
    \acro{SG}{Smart Grid}
    \acro{DT}{Digital Twin}
    \acro{CPS}{Cyber-Physical System}
    \acro{ICS}{Industrial Control System}
    \acro{IOT}{Internet-of-Things}
    \acro{MITM}{Man-in-the-Middle}
    \acro{HMI}{Human Machine Interface}
    \acro{BSS}{Battery Storage System}
    \acro{PV}{Photovoltaic}
    \acro{DOS}{Denial-of-Service}
    \acro{MV}{Middle Voltage}
    \acro{LV}{Low Voltage}
    \acro{IOA}{Information Object Address}
\end{acronym}

\bstctlcite{IEEEexample:BSTcontrol}

\title{Digital Twin for Evaluating Detective Countermeasures in Smart Grid Cybersecurity}

\author{
\IEEEauthorblockN{%
Ömer Sen\IEEEauthorrefmark{1}\IEEEauthorrefmark{2},
Nathalie Bleser\IEEEauthorrefmark{1},
Andreas Ulbig\IEEEauthorrefmark{1}\IEEEauthorrefmark{2},
}

\IEEEauthorblockA{%
\IEEEauthorrefmark{1}\textit{IAEW, RWTH Aachen Univserity,} Aachen, Germany\\
Email: \{o.sen, a.ulbig\}@iaew.rwth-aachen.de}
\IEEEauthorblockA{%
\IEEEauthorrefmark{2}\textit{FIT, Fraunhofer} Aachen, Germany\\
Email: \{oemer.sen, andreas.ulbig\}@fit.fraunhofer.de}
}

%

\maketitle

\IEEEpubidadjcol

\begin{abstract}

As the integration of digital technologies and communication systems continues within distribution grids, new avenues emerge to tackle energy transition challenges.
Nevertheless, this deeper technological immersion amplifies the necessity for resilience against threats, encompassing both systemic outages and targeted cyberattacks.
To ensure the robustness and safeguarding of vital infrastructure, a thorough examination of potential smart grid vulnerabilities and subsequent countermeasure development is essential.
This study delves into the potential of digital twins, replicating a smart grid's cyber-physical laboratory environment, thereby enabling focused cybersecurity assessments.
Merging the nuances of communication network emulation and power network simulation, we introduce a flexible, comprehensive digital twin model equipped for hardware-in-the-loop evaluations.
Through this innovative framework, we not only verify and refine security countermeasures but also underscore their role in maintaining grid stability and trustworthiness.
\end{abstract}



\section{Introduction} \label{sec:introduction}
Digitalization, particularly in the form of \ac{ICT}, has played a significant role in the modernization of distribution grids.
It enables improved grid management through real-time monitoring and control of assets, as well as the utilization of consumer flexibility resources and distributed generation in a market- and grid-aware manner.
However, as distribution grids become more reliant on \ac{ICT}, addressing the cybersecurity challenges associated with it is crucial~\cite{23_albarakati2018openstack}.
Cyberattacks pose a significant risk to the security and reliability of the grid, which could potentially disrupt the power supply~\cite{stuxnet_2011}.
Therefore, it is imperative to take proactive measures to prevent cyberattacks and mitigate the increasing cybersecurity risks~\cite{23_albarakati2018openstack}.
However, obtaining data on attack patterns in communication networks is challenging because such information is not readily available to the general public~\cite{3_zuech2015intrusion}.
To address these challenges and generate data in a flexible and scalable manner, virtualization of the grid has emerged as an intriguing approach~\cite{van2021towards}.
By creating a \ac{DT} of the grid infrastructure within a secure, isolated, and controllable environment, it becomes possible to closely replicate the grid's behavior~\cite{atalay2020digital}.
This replication involves various aspects, including operational management concepts, communication networks, asset behavior, and the physical plausibility of underlying processes.
Moreover, replicating the cybersecurity landscape is essential, ensuring that cyber-threats can be realistically reproduced without compromising the physical counterpart~\cite{sen2022investigating}.
This includes capturing the multi-stage behavior of cyberattacks, their interaction with assets and network services, and the propagation of traces within the communication network.
The challenges addressed in this work include:
\begin{enumerate}
\item Replication of the physical counterpart, encompassing grid operations, network communications, and the plausible behavior of assets and physical processes.
\item Replication of cyber-threats with realistic traces and plausible propagation behavior without impacting the physical counterpart.
\item Data generation for evaluating detective countermeasures without compromising their performance.
\end{enumerate}
In this paper, we investigate the suitability of simulation environments, especially \acp{DT}, for replicating \acp{CPS} such as \acp{SG} to evaluate the performance of detective countermeasures like \acp{IDS}.
We use a co-simulation framework, previously presented in works~\cite{van2021towards, sen2021approach, sen2022investigating}, to replicate a \ac{CPS} laboratory environment of a \ac{SG}, considering its process, communication, and operational aspects.
Using this framework, we execute cyberattack scenarios in both the laboratory and the \ac{DT}, generate data, and input it into selected \acp{IDS} to compare their detection quality.
We then utilize this comparison to assess the suitability of \acp{DT} for cybersecurity purposes, specifically for evaluating \acp{IDS}.
The contributions of this paper are as follows:
\begin{enumerate}
\item We present the simulation methodology used to create the \ac{DT} to replicate a \ac{CPS} reference, considering its process, communication, and operational aspects.
\item We discuss the methodology for replicating tailored cyberattacks in both the \ac{DT} and the laboratory environment for a comparative study.
\item We examine the suitability of \acp{DT} for evaluating \acp{IDS} in \acp{SG} through selected case studies.
\end{enumerate}
The remainder of this paper is structured as follows:
Section~\ref{sec:Background} presents the background analysis.
Section~\ref{sec:cosim} describes the design and approach.
Section~\ref{sec:investigation} discusses the investigation of the approach's suitability.
Lastly, Section~\ref{sec:conclusion} provides the conclusion.


\section{Analysis} \label{sec:Background}
In this section, we analyze the requirements, discuss the research landscape, and provide a problem analysis to define the scope of this research (Sections~\ref{subsec:Background_back}, \ref{subsec:Background_rw}, and \ref{subsec:Background_ps}).
\vspace{-0.25em}
\subsection{Requirement Analysis} \label{subsec:Background_back}
The structure of \acp{SG} and the considerations for the co-simulation approach are presented in this subsection.
\begin{figure}
\centering
\includegraphics[width=\linewidth]{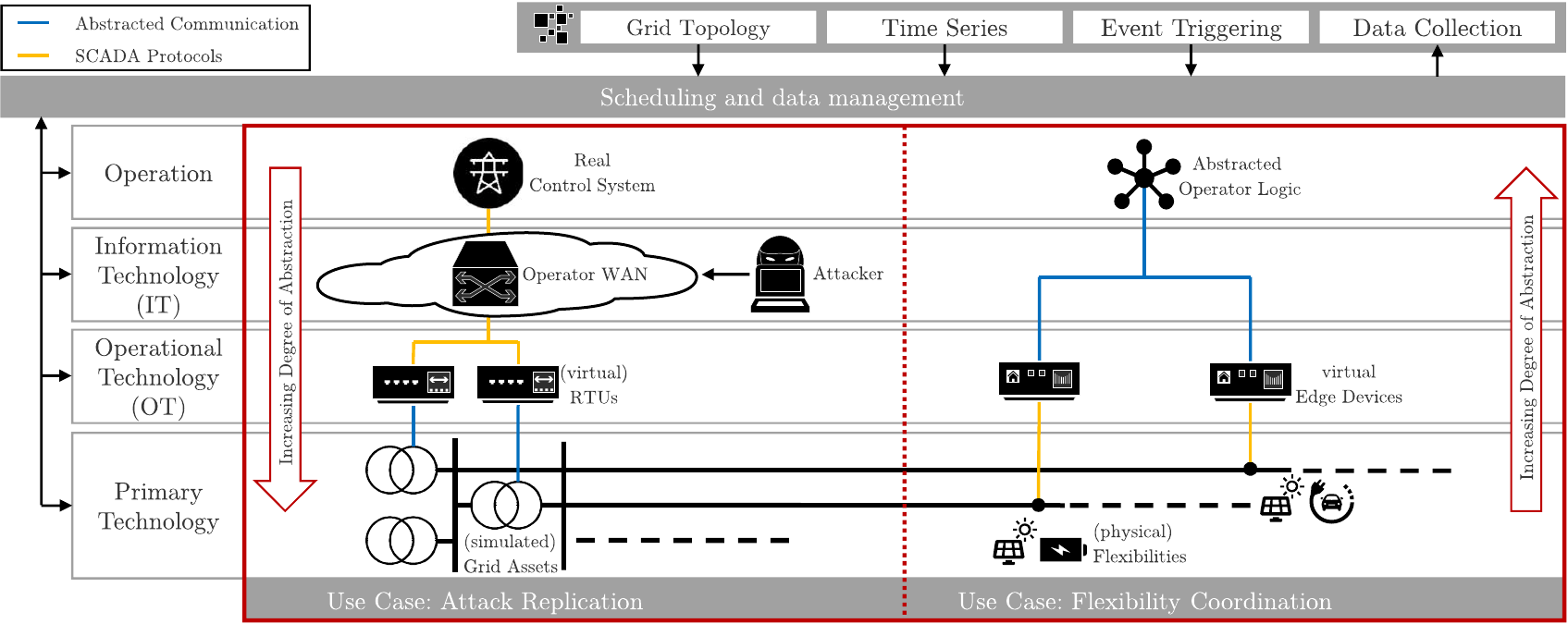}
\caption{Overview of the utilized co-simulation environment and its major components.
Depending on the use case, the respective components can be modeled in different degrees of abstraction.}
\label{fig:cosim}
\vspace{-1.5em}
\end{figure}
The overall structure of the \ac{SG} infrastructure, especially for the \ac{SCADA} use case, adheres to the Purdue model~\cite{williams1998purdue}.
It integrates the \ac{ET}, \ac{OT}, and \ac{IT} domains within a unified simulation environment.
The \ac{ET} domain encompasses primary equipment for energy distribution and transport, whereas the \ac{OT} domain is responsible for protection, control, and measurement through \acp{IED}.
Data from \acp{IED} is aggregated by \acp{RTU} and relayed to the \ac{SCADA} system via the \ac{OT} network using IEC 60870-5-104~\cite{IEC104_2006} or Modbus protocols~\cite{Modbus_2012}.
The \ac{SCADA} system communicates with remote sites via MTUs, and VPN connections facilitate remote access.
To depict the \ac{SG} as a \ac{CPS}, a consolidated framework is essential for charting interactions and data exchanges between domains.
The simulation environment must be both scalable and adaptable to account for variations in infrastructure characteristics.
For hardware-in-the-loop simulation and an accurate reproduction of network events, an event-based emulation of the communication network synchronized with discrete-time stepped simulation of power grids is proposed~\cite{mosaik_2019}.
This methodology enables the replication of grid operator reactions to changes and the examination of the repercussions of cyberattacks.
The co-simulation design should consider interactions across domains, facilitate the integration of additional simulators, and ensure data consistency.
Both \ac{IT} and \ac{OT} networks should be represented within the co-simulation framework to reproduce various cyber-threat scenarios.
Emulation of communication networks is favored for capturing dynamic cyberattacks, while simulation facilitates the expedited exploration of scenarios.
\vspace{-0.25em}
\subsection{Related Work} \label{subsec:Background_rw}
This subsection elaborates on methods and frameworks for replicating \ac{SG} systems and delving into cybersecurity through co-simulation.
Various methods, such as agent-based simulation and software-in-the-loop techniques, are employed for co-simulation~\cite{interfacing_ictpower_2018, cosim_survey_2018}.
Co-simulation frameworks chart cyberattacks and interruptions to analyze vulnerabilities in \acp{SG}~\cite{astoria_2016}.
Digital twinning, which combines virtualization and simulation, is utilized to study intricate systems, including applications in smart homes~\cite{qian2022digital, olivares2021towards}.
Research into cybersecurity in \ac{DT} extends to \ac{IOT} device integration and intrusion detection~\cite{atalay2020digital}.
Emulation, simulation, and hybrid testbeds are leveraged to appraise cyberattacks and craft security blueprints for \acp{SG}~\cite{oyewumi2019isaac, noorizadeh2021cyber, hammad2019implementation}.
The efficacy of intrusion detection is gauged through the replication of cyberattack scenarios.
Cybersecurity studies in \acp{SG} span co-simulation, mathematical modeling, and cyberattack trees to formulate sturdy security protocols.
\vspace{-0.25em}
\subsection{Problem Analysis} \label{subsec:Background_ps}
Comprehending cyber-threats and evaluating countermeasures in \acp{SG} pose challenges due to constrained resources and the imperative for a systematic approach.
Grid operators emphasize resilience and redundancy.
However, safeguarding other \ac{SG} domains demands supplementary resources, such as cyber incident data.
\ac{DT} and \ac{CPS} testbeds present potential solutions for probing into cybersecurity without jeopardizing the real grid reference, but they might affect the quality of output data.
If an adversary compromises the \ac{DT}, it could lead to unauthorized data access, manipulated simulations, security vulnerabilities in the real-world system, intellectual property theft, and a compromised blueprint for more effective attacks on the actual infrastructure.
Fully physically operated testbeds are expensive, and intricate simulations may lead to performance bottlenecks.
A hybrid approach, melding emulation with streamlined simulation, strikes a balance between data quality, performance, and adaptability.
To assess cybersecurity countermeasures in active distribution grids, a systematic approach is indispensable to reproduce threats and analyze inter-domain interactions.
This research formulates a co-simulation-based \ac{DT} methodology to replicate \ac{SG} segments and capture the operational sequence.
The emphasis lies on evaluating countermeasures in a laboratory setting as a prime target for cyber-threats.


\section{Digital Twin for Smart Grids} \label{sec:cosim}
In this section, we present the application of a \ac{DT} for investigating cybersecurity in smart grids, including an overview of the co-simulation environment, the \ac{DT} application, and cybersecurity considerations (Sections~\ref{subsec:cosim_overview},~\ref{subsec:cosim_dt}, and~\ref{subsec:cosim_dtsec}).
\vspace{-0.25em}
\subsection{Co-Simulation as Basis} \label{subsec:cosim_overview}
Based on the defined requirements discussed in Section~\ref{subsec:Background_back}, we have developed a co-simulation environment that combines the emulation of communication networks with the simulation of power grids~\cite{van2021towards}.
Figure~\ref{fig:cosim} provides an overview of the utilized co-simulation environment and its major components.
This environment can be integrated into various use cases and forms of interoperation, enabling \ac{DT} applications, custom simulation environments, or software- or hardware-in-the-loop integration into \ac{CPS}.
Different levels of abstraction can be modeled based on the use case, introducing variations in the simulation's detail.
The co-simulation design integrates steady-state simulation of power grids using \textit{pandapower}\cite{pandapower_2018} for real-time performance in hardware-in-the-loop simulation scenarios.
Emulation of communication networks is achieved using \textit{containernet}\cite{containernet_2016} or \textit{rettij}\cite{niehaus2022modern} with \textit{Docker} container simulators.
The co-simulation is centrally scheduled using a discrete time-based step coupling framework called Mosaik\cite{mosaik_2019}.
Mosaik allows the exchange of data between the power grid simulation and the communication network emulation, enabling a coherent co-simulation environment.
The use of communication network emulation provides a high level of detail required for cybersecurity assessment.
It allows for the simulation of \ac{OT} and \ac{IT} assets, such as \acp{IED}, \acp{RTU}, and \acp{MTU}, which communicate through industrial protocols like IEC 60870-5-104 and Modbus.
While there may be challenges in compatibility with accelerated simulation, the chosen small time step for simulation mitigates this issue for the co-simulation application at hand.
The data generated in this co-simulation environment, including raw communication data and synchronized time series data of the power system simulation, forms the basis for the \ac{DT} application and the design of the investigation environment.
\vspace{-1em}
\subsection{Digital Twin Application} \label{subsec:cosim_dt}
In this section, we detail the design of our \ac{DT} based on the co-simulation approach described in Section~\ref{subsec:cosim_overview}.
The objective of our \ac{DT} application is to investigate the performance of \ac{IDS} in both the \ac{DT} and a physical laboratory, and compare the detection quality in both systems.
%
The \ac{DT} serves as a replicated \ac{CPS} laboratory where selected case studies of cyberattack scenarios can be executed.
The system configuration remains the same across different scenarios, with variations in the attack scenarios.
During scenario execution, network traffic and power measurements are captured from both the \ac{DT} and the physical laboratory.
The resulting datasets are then fed into selected \ac{IDS} to generate alerts.
Finally, the datasets and alerts generated in the \ac{DT} and the physical laboratory are compared.
To emulate the behavior of the assets, we model the control logic of the corresponding \ac{OT} assets, such as \ac{MTU}, \ac{RTU}, and \ac{IED}.
Mapping of data points between the power grid simulation and industrial protocols, such as Modbus and IEC60870-5-104, is required.
The power grid simulation tracks the internal state of the power grid and its components, while also acting as an interface to the \ac{IED} components.
The \ac{MTU} serves as a communication interface for the operation management, and the \ac{RTU} acts as a gateway to the stations.
By implementing artificial flaws, bugs, and system vulnerabilities, the \ac{DT} can replicate various attack scenarios, such as \ac{DOS} or code manipulation attacks.
%
%
\vspace{-0.25em}
\subsection{Cybersecurity Considerations} \label{subsec:cosim_dtsec}
\begin{figure}
\centering
\includegraphics[width=0.8\linewidth]{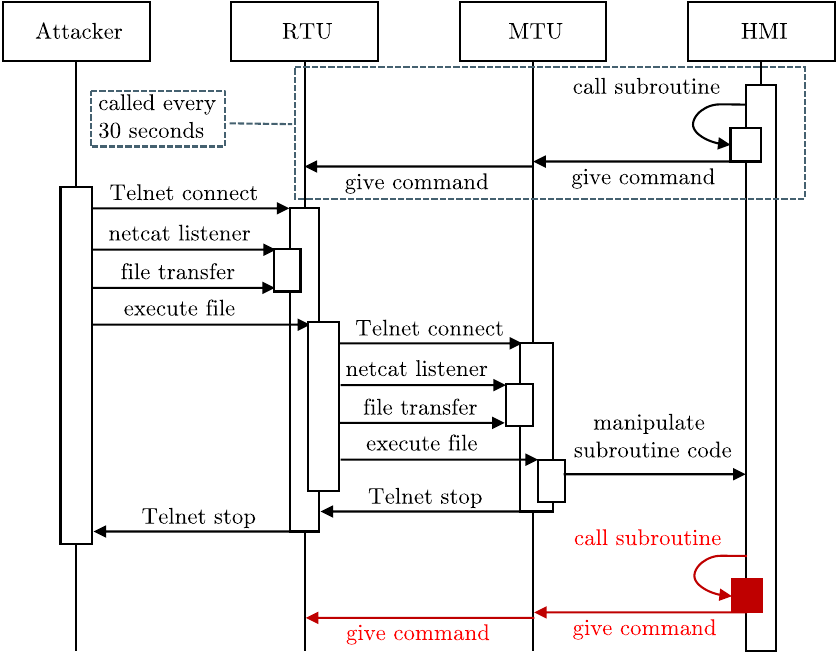}
\caption{Illustration of the 2-hop attack pattern in the \ac{DT} application.}
\label{fig:2hop}
\vspace{-1.5em}
\end{figure}

In this section, we explain the cybersecurity considerations involved in the \ac{DT} application, including attacker design and \ac{IDS} selection.
For our investigation purposes, the attacker needs to have the capabilities to spread across the network, infect hosts, make changes to infected devices, leave traces in network traffic, and have an impact on the grid.
The attacker's behavior is implemented through a set of capabilities, including establishing Telnet or SSH connections, executing arbitrary commands, transferring files, and performing network scans using NMAP.
Figure~\ref{fig:2hop} illustrates an attack pattern created using these capabilities.
The attacker's target is the \ac{HMI} component of the \ac{MTU}.
The \ac{HMI} is infected first and then used to infect the \ac{MTU}.
During the infection process, the attacker establishes a Telnet connection to the target, transfers the attack file, and performs the next steps of the attack on the target device.
This attack pattern showcases the potential spread of the attacker within the system.
Given the collected dataset of network traffic and active power, we require \acp{IDS} capable of detecting suspicious packets and outliers in active power and network traffic.
We selected three \acp{IDS} for our investigation: Snort, a signature-based \ac{IDS}; a specification-based \ac{IDS} presented in~\cite{sen2022specification}; and a statistical anomaly-based \ac{IDS}.
Snort uses rule sets to detect attack signatures in network traffic.
The specification-based \ac{IDS} utilizes internal state machine models to detect anomalies in IEC60870-5-104 traffic.
The statistical anomaly-based \ac{IDS} detects outliers in active power and network traffic based on statistical analysis.
By comparing the generated datasets and alerts between the \ac{DT} and the physical laboratory, we can evaluate the effectiveness of the selected \acp{IDS} in both systems and analyze the differences in detection quality.
\vspace{-0.25em}
\subsection{Attack Implementation}
\begin{figure}
	\centering
	\includegraphics[width=0.8\linewidth]{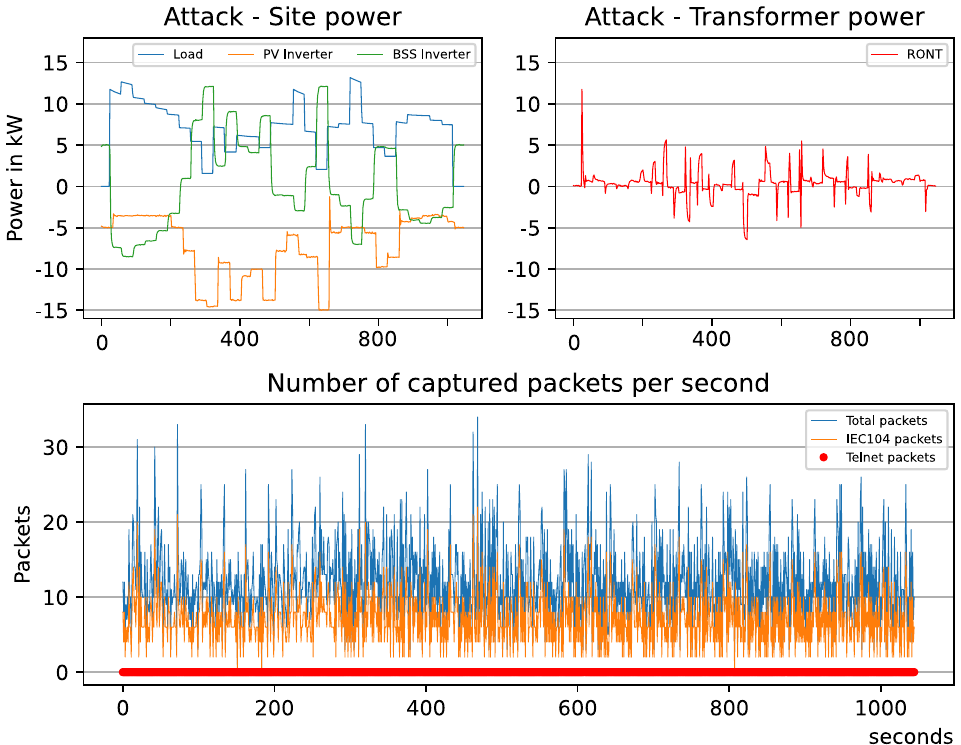}
	\caption{Behavior when not under attack}
	\label{fig:no_attack}
\vspace{-1.5em}
\end{figure}

\begin{figure}
	\centering
	\includegraphics[width=0.8\linewidth]{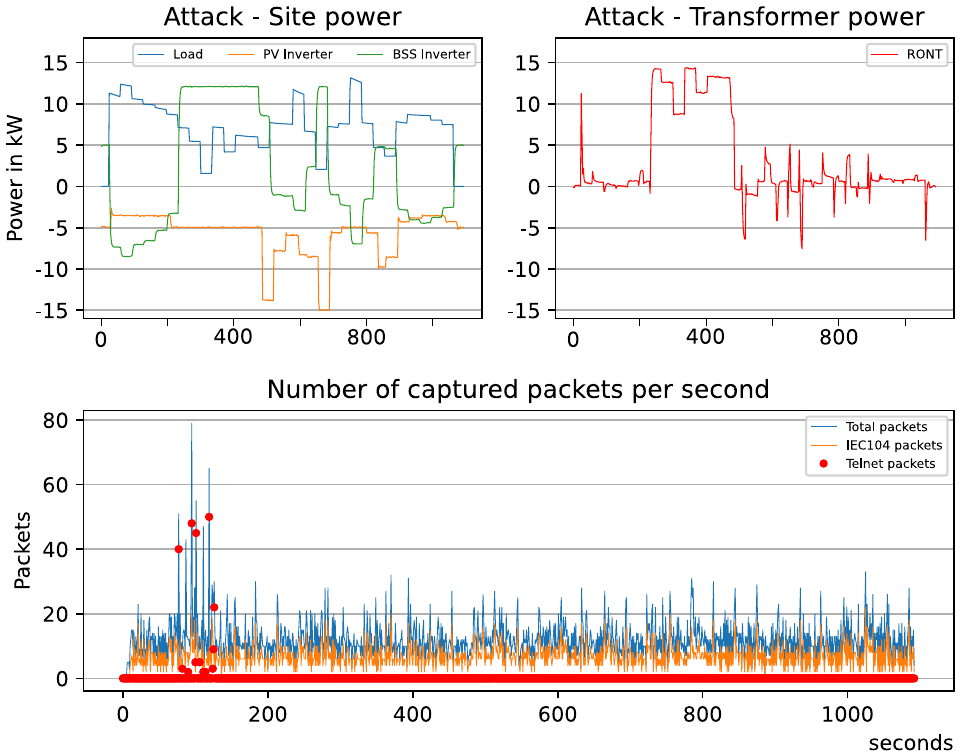}
	\caption{Results of targeting true power manipulation}
	\label{fig:result_manipulation}
\vspace{-1.5em}
\end{figure}

\begin{figure}[t]
	\centering
	\includegraphics[width=0.8\linewidth]{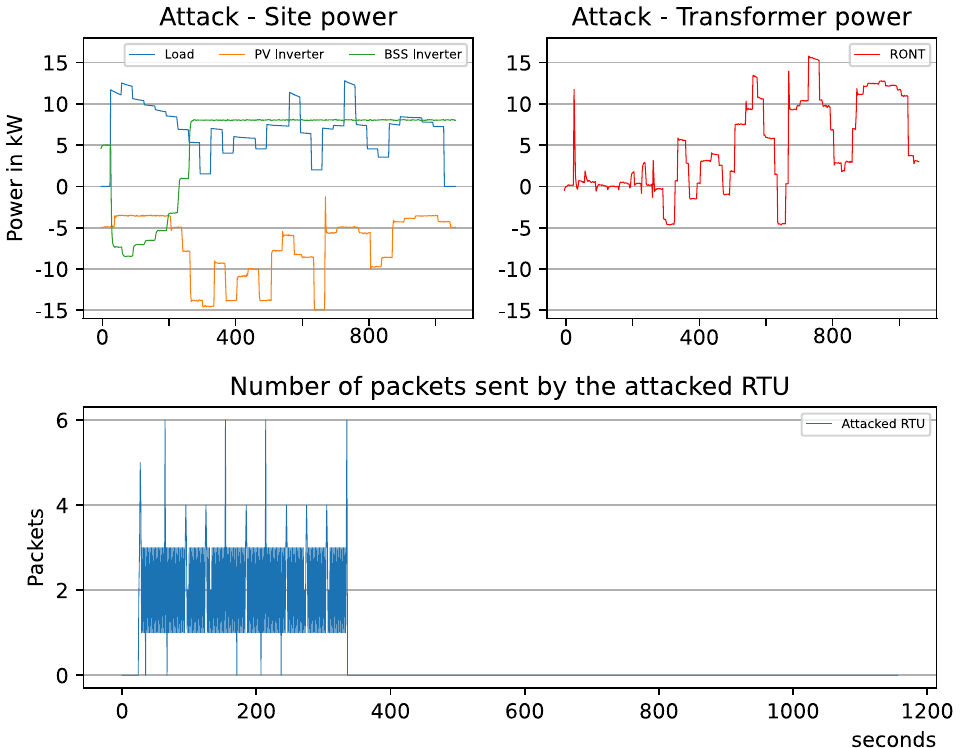}
	\caption{Results of targeting \ac{RTU} shutdown}
	\label{fig:result_shutdown}
\vspace{-1.5em}
\end{figure}

\begin{figure}[t]
	\centering
	\includegraphics[width=0.8\linewidth]{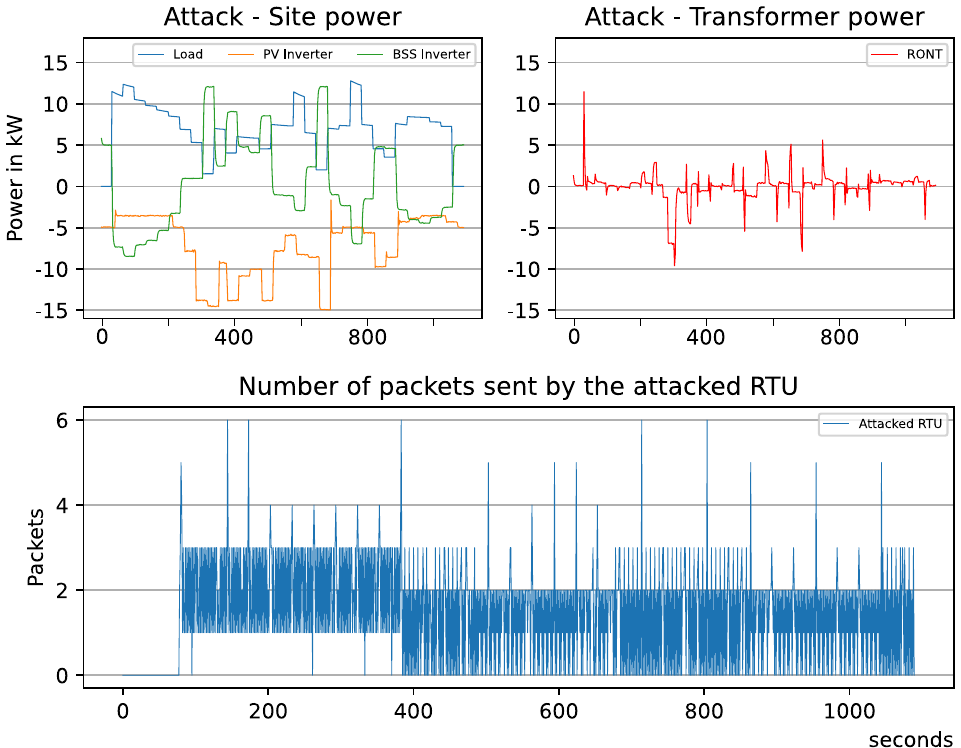}
	\caption{Results of targeting \ac{RTU} slowdown}
	\label{fig:result_slowdown}
\vspace{-1.5em}
\end{figure}

The attacks are programmed and tested on the \ac{DT} before being ported to the laboratory.
Porting the attacks to the laboratory posed minimal challenges, primarily involving adapting to differences in file structure, usernames, and passwords.
To ensure the safety of real hardware, the \ac{HMI}, \ac{MTU}, and \ac{vRTU} code running on docker containers are installed on Raspberry Pis.
Only one \ac{RTU} is replaced with the \ac{vRTU} code, while the other \acp{RTU} remain connected as usual.
The IP addresses of the replaced \acp{RTU} are blacklisted in the attack code to prevent any harm.
Simulating cyberattacks on the power grid involves identifying attack vectors like the \ac{MITM} attack, which intercepts and manipulates communications between \ac{MTU} and \acp{RTU}, highlighting vulnerabilities and the potential for undetected command manipulations.
Another method involves attackers manipulating update files at the operating level, or breaching the internal network using default passwords, enabling them to control devices or access critical data.
During the experiment, the true power is varied every 30 seconds, as depicted in Figure~\ref{fig:no_attack}.
Traffic from the switch to the \ac{MTU} and from the attacker to the switch is captured.
The baseline traffic without any attacks is visible in Figure~\ref{fig:no_attack}.
In the 2-hop attack, the attacker initiates the attack from a device within the process network.
After conducting reconnaissance, the attacker targets the \ac{RTU} by establishing a Telnet connection and transferring an executable Python file using netcat.
The same file transfer steps are then executed from the \ac{RTU} to the \ac{MTU}.
The attack file modifies the command-handling files on the \ac{MTU}.
Three attacks were conducted on the \ac{MTU}: true power manipulation, \ac{RTU} shutdown, and \ac{RTU} slowdown.
The results are presented in Figures~\ref{fig:result_manipulation},~\ref{fig:result_shutdown}, and~\ref{fig:result_slowdown}, respectively.
In the true power manipulation attack, the commands sent to the \ac{RTU}s are manipulated, resulting in spikes in transformer power load.
The Telnet connections and file transfers during the attack are visualized by the spike in the total packets sent and Telnet packets sent, as shown in Figure~\ref{fig:result_manipulation}.
The \ac{RTU} shutdown attack involves sending an unknown \ac{IOA} to the \ac{RTU}, causing the \ac{vRTU} to shut down.
Consequently, the \ac{BSS} Inverter stops responding to the \ac{MTU}, leading to spikes in transformer power.
The targeted \ac{RTU} also ceases sending packets, as depicted in Figure~\ref{fig:result_shutdown}.
In the \ac{RTU} slowdown attack, an unknown \ac{IOA} is sent to the \ac{RTU}, resulting in its slowdown.
This attack involves replacing a command with an incorrect \ac{IOA}, leading to a power drop.
Moreover, the slowdown reduces the number of packets sent per second.
Figure~\ref{fig:result_slowdown} provides a visualization of these effects.

\section{Investigation} \label{sec:investigation}
In the investigation, scenarios described in Section~\ref{subsec:scenarios} are executed five times in both the \ac{DT} and the laboratory, used for evaluation of three selected \ac{IDS} (Section~\ref{subsec:ids}) with discussion in Section~\ref{subsec:discussion}.

\vspace{-0.25em}
\subsection{Scenarios} \label{subsec:scenarios}
The investigation considers the following scenarios:
\begin{inparaenum}[(i)]
\item No attack,
\item 2-hop Telnet attack using one of three command manipulations,
\item Exfiltration attack using Telnet and the tail command, and
\item Reconnaissance attack using NMAP and attempted Telnet logins.
\end{inparaenum}
The laboratory used for the scenarios simulates cyberattacks on power grids using a medium- and low-voltage grid, integrated with equipment like controllable transformers, battery storage systems, and PV inverters, all connected to a control center via telecontrol devices.
This comprehensive setup, featuring a process data network with four ICT switches and a direct control room connection, facilitates in-depth cross-domain investigations of cyberattack impacts on power grids.
The scenario without attacks serves as a baseline to evaluate the impact of the attacks and distinguish false positive alerts from true positive alerts issued by the \ac{IDS}.
The same scenario used in previous work for normal operation~\cite{hacker2021framework} is employed.
The 2-hop Telnet attack primarily tests the capability of the specification-based \ac{IDS} to detect command tampering and the ability of the statistical \ac{IDS} to detect the impact on active power measurements.
The consequences of disturbances in the grid considered in this scenario include a slowdown of the \ac{HMI}, resulting in fewer transmitted packets, a shutdown of the \ac{HMI}, causing wild spikes and a drop in active power on the \ac{LV} busbar of the secondary substation, and tampered values within commands aimed at creating power spikes that persist for several 30-second intervals.
The exfiltration attack simulates the extraction of critical data from either the \ac{HMI} or \ac{MTU}.
The impact on network traffic should be observable by all \ac{IDS}, but there should be no noticeable impact on system performance.
The Recon attack tests the Snort \ac{IDS}' ability to detect NMAP scans and failed Telnet logins, as well as the statistical \ac{IDS}' capability to detect associated spikes in total bytes per second.
The specification-based \ac{IDS} can only detect Telnet packets transmitted during logins.
\vspace{-0.25em}
\subsection{Intrusion Detection Results} \label{subsec:ids}
\begin{figure}
\centering
\includegraphics[width=0.95\linewidth]{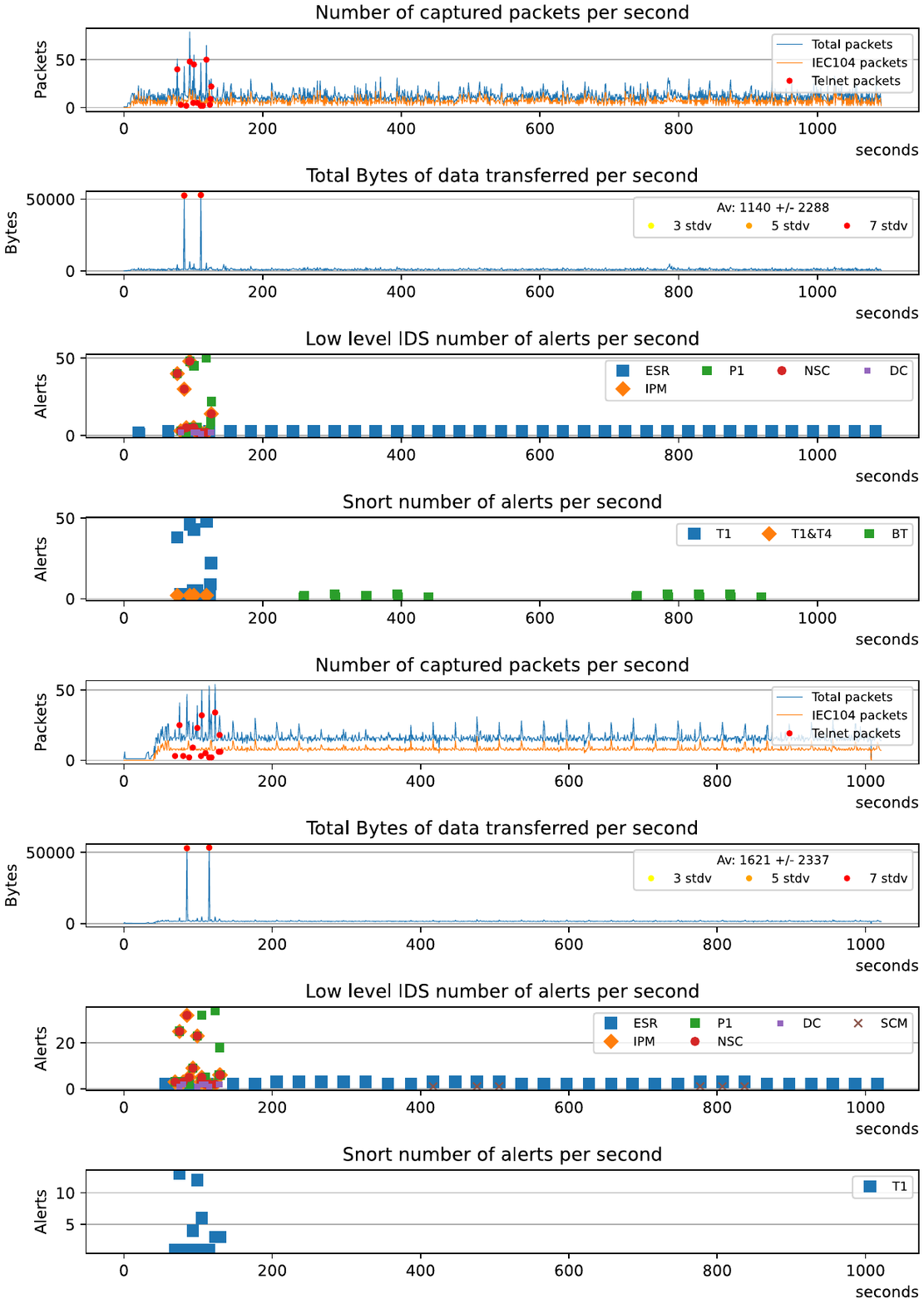}
\caption{Example results of traffic analysis by \ac{IDS}.
The first four diagrams are related to the laboratory, while the last four diagrams depict the \ac{DT}.}
\label{fig:ids}
\vspace{-1.5em}
\end{figure}

Figure~\ref{fig:ids} presents an example result of traffic analysis by the \ac{IDS} in the 2-hop Telnet scenario.
The first and fifth diagrams illustrate the type of traffic transmitted during the scenario, while the other diagrams provide insights into the alerts generated by the \ac{IDS}.
The statistical \ac{IDS} applied to the total number of bytes per second functions similarly in both the \ac{DT} and the laboratory.
As depicted in the second and sixth diagrams of Figure~\ref{fig:ids}, all spikes in the total number of bytes are detected and classified as particularly severe.
The specification-based \ac{IDS} performs consistently in both the \ac{DT} and the laboratory.
As shown in the third and seventh diagrams of Figure~\ref{fig:ids}, it correctly detects all Telnet packets (P1 alert), identifies suspicious keywords in the Telnet packets (DC alert), and raises alerts for unknown IP addresses (IPM alert), unknown connections (NSC alert), and unknown IOAs.
However, it generates false alarms (ISR alarm) for regular sets of \ac{MTU} commands, even without an attack, suggesting the need for adaptations in the underlying state machine.
The Sequence Control Mismatch warning is primarily triggered in the \ac{DT} records, indicating slightly different traffic behavior between the \ac{DT} and the laboratory.
The Snort-Wireshark dissector produces mixed results for the \ac{DT} and the laboratory.
The outcomes can be observed in the fourth and final graph of Figure~\ref{fig:ids}.
It successfully detects Telnet packets (T1 and T4 alerts), but while it accurately detects all Telnet packets in the laboratory, it does not perform as well in the \ac{DT}.
Additionally, Snort triggers alerts about bad traffic in the laboratory record, which is not a false alarm but a consequence of the way traffic was captured in that environment.
\begin{figure}
    \centering
    \includegraphics[width=0.90\linewidth]{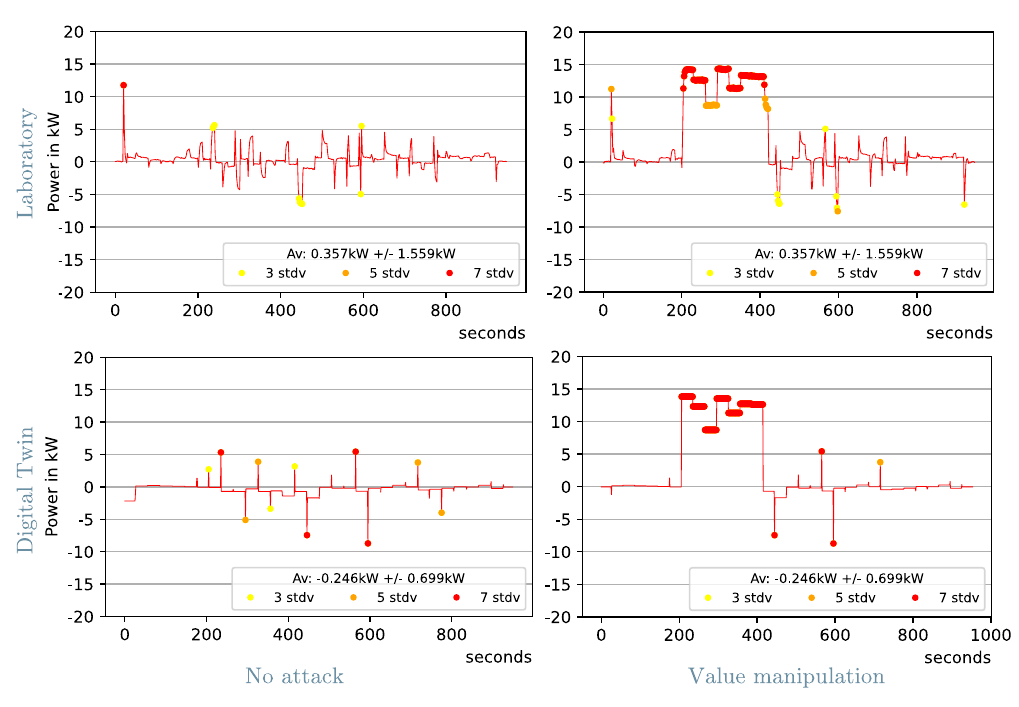}
    \caption{Example result of the statistical \ac{IDS}.}
    \label{fig:power_ids}
\vspace{-1.5em}
\end{figure}
Figure~\ref{fig:power_ids} illustrates an example result of the statistical \ac{IDS} for the active power on the \ac{LV} busbar of the secondary substation in the "no attack" scenario and the command manipulation scenario.
The severity of the detected spikes is represented by colored dots.
The left side of the figure displays the change in active power during the "no attack" scenario, while the right side shows the variation in active power during the attack with command manipulation.
It can be observed that some parts of the attack are marked orange (medium severity) in the laboratory diagrams but red in the \ac{DT} diagrams.
However, slight differences were observed in the active power on the \ac{LV} busbar of the secondary substation between the two datasets.
These differences are attributed to transients and variations in the response times of the assets in the laboratory, resulting in a larger standard deviation compared to the \ac{DT}.
The \ac{DT} does not consider these transients or variations in the response time of the devices, leading to a smaller standard deviation.
As a result, the statistical \ac{IDS} classifies all attack effects on the \ac{DT} as severe, while some impacts are classified as low.
\vspace{-0.25em}
\subsection{Discussion} \label{subsec:discussion}
Based on the obtained results and findings, an overall good performance can be concluded.
Various \ac{IDS} were evaluated in the experiments, including (i) signature-based, (ii) specification-based, and (iii) statistical-based \ac{IDS} on (iv) Bytes transferred per second and (v) active power.
The first two \ac{IDS} were designed to detect suspicious packets, but only the specification-based \ac{IDS} performed equally well on both laboratory and \ac{DT} traffic, while the signature-based \ac{IDS} performed poorly in both scenarios.
The third \ac{IDS} aimed to identify outliers in Bytes transferred per second and performed well in both scenarios.
The fourth \ac{IDS} aimed to detect outliers in power load and effectively detected all attacks that affected the power load, but the attacks on the laboratory power load were considered less severe.
\acp{DT}, when replicating intricate systems like power grids for cybersecurity evaluations, face challenges in capturing system complexities, ensuring real-time high-fidelity simulations, and adapting to evolving grid dynamics.
Additionally, the security of the twin itself is crucial, as its compromise could provide misleading results or expose real system vulnerabilities.
Overall, evaluating \ac{IDS} on the \ac{DT} proved to be a good alternative to evaluating them in the laboratory.
The detection quality based on the data generated by the \ac{DT} closely matched the performance of the physical counterpart in most scenarios.
Additionally, the \ac{DT} provided a suitable environment for developing attacks that could be used in both systems, as porting the attacks from the \ac{DT} to the laboratory was seamless, and the attacks performed similarly in both environments.
This capability can also be utilized to generate training data for machine learning-based \ac{IDS}.


\section{Conclusion} \label{sec:conclusion}
In this paper, we have investigated the suitability of using a \ac{DT} for evaluating \ac{IDS} in the context of \ac{SG}.
Our \ac{DT} approach is based on a modular co-simulation framework that can be customized to simulate \ac{SG} across all its domains.
This flexibility allows us to adjust the abstraction level of the components specifically for our \ac{DT} application.
Furthermore, we utilized a \ac{CPS} laboratory environment as a reference system to compare the detection quality of the tested \acp{IDS}.
By simulating identical attack scenarios with varying impacts on the network and grid in both the \ac{DT} and the laboratory, we could observe that \ac{DT} provides a valuable environment for evaluating detective countermeasures in \ac{SG}.
In future work, we plan to further validate our findings by incorporating more scenarios and \acp{IDS}.
Additionally, we aim to leverage the \ac{DT} to generate attack data that can be used for training machine learning-based \acp{IDS}.
This approach can be implemented in a manner similar to our investigation setup, allowing for cross-validation of the results with our \ac{CPS} laboratory.


\vspace{-0.5em}
\bibliographystyle{IEEEtran}
\bibliography{cas-refs}

\end{document}